\documentstyle[aps,eqsecnum,preprint]{revtex}

\begin{document}

\draft

\title{Activated resistivities in the integer quantum Hall effect}
\author{Sudhansu S. Mandal \cite{e1} and V. Ravishankar
\cite{e2}}
\address{Department of Physics, Indian Institute of technology, 
Kanpur -- 208 016, INDIA }

\maketitle

\begin{abstract}
We have determined the off-diagonal and diagonal conductivities for a
quantum Hall effect system at exactly integer filling at
finite temperatures and in the presence of weak short ranged disorder
potential within the self
consistent Born approximation. We find that there is a finite
temperature contribution to off-diagonal conductivity $\sigma_{xy}$
which is
`anomalous' in nature as it survives even in the zero impurity
limit. The diagonal conductivity $\sigma_{xx}$
survives only when both temperature
and disorder is non zero. At low temperatures, $\sigma_{xx}$
activates with a
temperature dependent prefactor. Inverting the conductivity matrix,
we determine the resistivities. The deviation of the off-diagonal
resistivity $\rho_{xy}$
from its zero temperature value and the diagonal
resistivity $\rho_{xx}$
activate with a temperature dependent prefactor at low
temperatures, in agreement with experiments. Further, we 
find two physical regimes both of which are at low
temperatures and low broadening, which provide the experimentally
observed linear relationship between the deviation of $\rho_{xy}$
and the $\rho_{xx}$ with different signs. We 
have also estimated the effective masses from the experimental data of
$\rho_{xy}$ and find them to be reasonable. Finally, our result
on compressibility as a function of temperature shows that there is
no phase transition involved in the system as far as the
temperature is concerned.

\end{abstract}



\section{INTRODUCTION}

It was first observed by von Klitzing et al. \cite{klit} that
two dimensional electron gas at low temperatures and in the
presence of high magnetic field $B$ perpendicular to the plane 
can exhibit
quantization of the Hall resistivity at integer fillings,
with a high accuracy (one part in $10^5$). 
Subsequently, it has been observed that the precision of this
quantization could be very high (one part in $10^9$)
\cite{expt1}. The essence of this integer quantum Hall effect
(IQHE) is that the quantization of Hall resistivity $\rho_{xy}
=\frac{1}{i} (2\pi /e^2)$ (we have set the unit $\hbar =c =1$) at
integer filling factor $\nu =i$ exists for a wide range of physical
parameters, viz, of $B$ and of the carrier density $\rho$.
At the same time the diagonal resistivity $\rho_{xx}$ shows a sharp
minimum. Prange \cite{prange}, Laughlin \cite{laugh}, and Halperin
\cite{halp} have argued that as long as the Fermi level lies in the
region of localized states between two current carrying regions of
extended states, the Hall conductivity $\sigma_{xy}$ is quantized
and $\sigma_{xx}$ vanishes. It has been observed that the measured value of
$\rho_{xy}$ approaches the universal value $\frac{1}{i}(2\pi /e^2)$
as the temperature is lowered.

In his pioneering work, Laughlin \cite{laugh} has shown that the
edge effects are not important for the accuracy of quantization. He
further speculates that the only significant source of error in
quantization is the thermal activation. In fact, there exists a
fairly good number of experiments
\cite{yoshi,cage1,wel,weiss,clark,kata,wei1,wei2,huck1,hwang,waka,koch}
which study the effect of temperature on the quantization. Speaking
broadly, there 
are two important aspects regarding these studies, viz, (i) the
flatness of the plateaus formed in $\rho_{xy}$ and the critical
transition between the plateaus, (ii) the value of quantization at
the central point of the plateaus, i.e., the points at which
$\rho_{xx}$ show minima and for which $\nu$ are exactly integers.

The present paper is principally concerned with the latter aspect,
i.e., the behaviour of $\rho_{xy}$ and $\rho_{xx}$ with changing
temperature and disorder. We shall study this at the centre of the
plateau, i.e., we look at the point at which the filling factor
$\nu$ is an integer.

Let us now briefly review the pertinent experimental situation.
Early studies by Yoshihiro et al \cite{yoshi} and Cage et al
\cite{cage1} show that $\rho_{xy}$ decreases with increase in
temperature at the minimum of $\rho_{xx}$. They also find a
linear relationship $\Delta \rho_{xy} =-S\rho_{xx}^{{\rm min}}$
between the deviation $\Delta \rho_{xy} \equiv \rho_{xy} (T) -\rho_{xy}
(0)$ from the zero temperature value of $\rho_{xy}$, and the minimum
value of $\rho_{xx}$. Cage et al \cite{cage1} have further found that the
value of $S$ varies from 0.06 to 0.51 for different GaAs samples. In
Si MOSFETs, Yoshihiro et al \cite{yoshi} have found $S \simeq 0.1$.
Although the values of $S$ are device dependent 
and also on how the system
is cooled, the
linear relationship is itself universal.
Cage \cite{cage2} has pointed out that
sample size effects which have been calculated by Rendell and
Girvin \cite{rend}, Hall probe misallignment, and variable range
hopping conduction are not responsible for the linear
relation between $\Delta \rho_{xy}$ and $\rho_{xx}^{{\rm min}}$.
Moreover, Cage et al \cite{cage1} have surprisingly observed that
even when the Hall steps are flat to within 0.01 ppm instrumental
resolution, the temperature dependent error $\Delta \rho_{xy}$ is
still quite large. This suggests the loss of universality at finite
temperatures. However, it is not always true that $\rho_{xy}$
decreases with the increase of temperature, as Wel et at \cite{wel}
have observed a positive slope between $\Delta \rho_{xy} $ and
$\rho_{xx}^{{\rm min}}$ in their GaAs samples. Finally,
Weiss et al \cite{weiss} have fitted the temperature dependent
prefactor in the activation of $\rho_{xx}^{{\rm min}}$ by a form
$\rho_{xx}^{{\rm min}} \sim \frac{1}{T} \exp [ -\omega_c /2T ]$ in
the leading order, where $\omega_c$ is the cyclotron frequency.

We remark here the fact 
that the value of $S$, in general, depends not only
on $T$ but also on its prior history, hinders
any explicit comparison between theory and experiment. Nevertheless,
we show that the present model 
yields a linear relationship between $\Delta \rho_{xy}$ and
$\rho_{xx}^{{\rm min}}$ and can accommodate both positive and 
negative values of $S$. The linearity is in a smaller
range of temperatures for the case of positive values of $S$.
A further estimation of effective mass $m^\ast$ 
which follows thereof is also not
unrealistic, and is of the same order of magnitude as the
experimental number.

The experimental results are quite ambiguous regarding the
prefactor of $\sigma_{xx}$. Clark et al \cite{clark} have measured
the prefactor of $\sigma_{xx}$ for fractional quantum Hall states
and find a universal value $e^2/2\pi $, independent even of the
filling factors. 
In contrast, in a recent experiment by Katayama et al 
\cite{kata}, the prefactor is found to be proportional to $1/T$.
On the theoretical side, Fogler and Shklovskii \cite{fogl} have
found that for the case of long range random potential, 
the prefactor is indeed 
universal and is given by $e^2 /2\pi $, but only
above a certain critical
temperature, and that it decays according to a power law below the
critical temperature. On the other hand, Polyakov and Shklovskii
\cite{poly} obtain the prefactor to be $2e^2 /2\pi $. 
Here we study
the case of short range potential and we find the prefactor to be 
temperature dependent, in agreement with that of Katayama et al
\cite{kata}.

To be sure, there is more experimental information available
largely related to the aspect (i) mentioned above and not studied
in the present paper. For instance, Wei et al \cite{wei1} have
observed a similarity between $\rho_{xx}$ and
$\frac{d\rho_{xy}}{dB}$ with the only difference that while the
maximum value of $\rho_{xx}$ decreases, the maximum value of
$\frac{d\rho_{xy}}{dB}$ increases with decreasing temperature over
the range of temperatures from 0.1 K to 4.2 K. They have also found 
the power law behaviour $(\frac{d\rho_{xy}}{dB})^{{\rm max}}
\propto T^{-k}$ and the width of the $\rho_{xx}$ peak $\Delta B
\propto T^{k}$ with $k=0.42 \pm 0.04$. Further measurements by Wei
et al \cite{wei2} show that the extrema of $\frac{d^2 \rho_{xy}}{
dB^2}$ and $\frac{d^3\rho_{xy}}{dB^3}$ diverge like $T^{-2k}$
and $T^{-3k}$ respectively. 
Huckestein et al \cite{huck1}
also have measured the temperature dependence of the plateau 
and obtain a value $k =0.42$ in agreement with Wei et al \cite{wei1}.
The above observations are for fully polarized quantum Hall
states. When the Landau levels are spin degenerate, Wei et al
\cite{wei2} and Hwang et al \cite{hwang} have reported that 
$(\frac{d\rho_{xy}}{dB})^{{\rm max}}$ and $(\Delta B)^{-1}$ 
diverge like $T^{-k/2}$. However, Wakabayashi \cite{waka}
and Koch et al \cite{koch} find no evidence for the universality
of the exponent $k$. All these aspects are discussed in a recent
review by Huckestein \cite{huck2}.
These mutual conflicting experimental results lie outside the scope of
the present paper.

On the theoretical side, Ando et al \cite{ando} 
computed $\sigma_{xx}$
and $\sigma_{xy}$ at $T=0$ using a simple 
Lorentzian density of states (DOS)
approximated from the self consistent Born approximation (SCBA)
density of states; they have not considered the
frequency dependent imaginary part of the SCBA self energy of the
single particle Green function. In this paper, we make the full
use of frequency dependent self energy.

Finally, there is yet another novel aspect that emerges from the present
study, viz., the temperature evolution of $\sigma_{xy}$, even for a
pure system! Recall the classical argument \cite{prange} that the
translationally invariant system does not lead to any temperature
dependence on $\sigma_{xy}$. Therefore the Maxwell gauge
interactions that are at play here belie such a naive expectation.

The plan of the paper is as follows. In the next section we have
discussed the formalism. In section III, single particle Green's
function is determined within the SCBA. In section IV, we have
evaluated response function and subsequently the off-diagonal and
diagonal conductivities. Section V is devoted for determining the
resistivities and comparison with experiments. We conclude the
paper in section VI. Finally, we have computed the compressibility
for the integer quantum Hall states in the appendix B.

\section{FORMALISM}

Consider a system of (weakly) interacting electrons in two space 
dimensions in the presence of a uniform external magnetic field of
strength $B$, confined to the direction perpendicular to the plane.
The electrons also experience a short 
ranged impurity potential $U(X)$.
The strength of the magnetic field 
is fine tuned such that $N$ Landau levels (LL) 
are exactly filled.
In the presence of sufficiently high magnetic
fields (as is relevant to our case),
the spins of the fermions would be `frozen' in the direction of
magnetic field. Therefore, one may treat the fermions as
spinless. The study of such a {\it spinless} system can be
accomplished with the 
Lagrangian density \cite{fnote1},
\begin{equation}
{\cal L} = \psi^\ast iD_0\psi -\frac{1}{2m^\ast} 
\vert D_k \psi \vert^2 +\psi^\ast \mu \psi -\psi^\ast U \psi -
eA_0^{\mbox{in}}\rho +\frac{1}{2}\int d^3x^\prime
A_0^{\mbox{in}} (x)V^{-1}(x-x^\prime)A_0^{\mbox{in}} (x^\prime) \, .
\label{eq1}
\end{equation}
Here $D_\nu =\partial_\nu -ie(A_\nu +A_0^{\mbox{in}}\delta_{\nu ,0} )$
(where $A_\nu $ is the external Maxwell gauge field and 
$A_0^{\mbox{in}}$ is identified as internal scalar potential),
$\mu$ is the chemical potential, and $m^\ast$
and $\rho$ are the effective mass and the mean density of electrons
respectively. The fifth term in Eq.(\ref{eq1}) describes the charge
neutrality of the system. Finally, $V^{-1} (x-x^\prime)$ 
represents the inverse of the instantaneous charge 
interaction potential (in the operator
sense). The above Lagrangian density is equivalent to the usual 
interaction term with quartic form of fermi fields,
which can be obtained by an integration of
$A_0^{\mbox{in}}$ field in Eq.~(\ref{eq1}).
This form of the Lagrangian is obtained by Hubbard-Stratonovich
transformation using an auxiliary field $A_0^{{\rm in}}$.
Note also that the electrons interact with each other
via $1/r$ or some other short range potential,
i.e., the internal dynamics
is governed by the (3+1)-dimensional Maxwell Lagrangian
as is appropriate for the medium.

We therefore construct the partition function ($\beta  =1/T$ being
the inverse temperature) of the system,
\begin{equation}
{\cal Z} = \int\, [dA_\tau^{\mbox{in}}][d\psi ][d\psi^\ast ] \exp \left[
-\int_0^\beta d\tau\,\int d^2X {\cal L}^{(E)} \right] \, ,
\label{eq2}
\end{equation}
which on integration over the fermionic fields, (by fixing of the saddle
point at the uniform background magnetic field $B$),
factors into the form ${\cal Z} ={\cal
Z}_B {\cal Z}_I$. Here ${\cal L}^{(E)}$ is the Euclidean version
of ${\cal L}$ in Eq.~(\ref{eq1}). For the transformation into the
Euclidean space, we make a substitution $t \rightarrow  -i\tau$ and
consider the real parameter $\tau $ as a coordinate on a circle of
circumference $\beta$. Fermionic fields are antiperiodic on this
circle while the bosonic fields are periodic. The time component of
the vector field $A_\mu$ is redefined as $A_0 \rightarrow
iA_\tau$.
The back ground part of the partition function is given by
\begin{equation}
{\cal Z}_B = Tr \, e^{-\beta (H-\mu)} \; ,
\label{eq3}
\end{equation}
which is obtained from the fermion determinant, found by the
integration of fermionic fields. The finite temperature back ground
properties of the system can be studied using ${\cal Z}_B$. Here
\begin{equation}
H = -\frac{1}{2m^\ast} D_k^2 +U(X) \equiv H_0 +U(X)
\label{eq4}
\end{equation}
is the single particle Hamiltonian. 
${\cal Z}_I$ is the partition function corresponding to 
the external probe, which will be determined later below.

\section{SINGLE PARTICLE GREEN'S FUNCTION}

\subsection{Disorder free Green's function}

In the absence of disorder, $U(X)=0$, the spectrum of $H$ for the
system is a set of Landau levels (LL) with energy eigen value for
the $n$-th LL,
\begin{equation}
\epsilon_n = (n+\frac{1}{2})\omega_c  \; ;\; n=0,1,2,\cdots \; , 
\label{eq2.1}
\end{equation}
where $\omega_c =eB/m^\ast $ is the cyclotron frequency. In the
Landau gauge
\begin{equation}
\vec{A} = (-BX_2 , 0) \; , 
\label{eq2.2}
\end{equation} 
the eigen function corresponding to the eigen value $ \epsilon_n$ is
\begin{equation}
\psi_{nk} (X) =\frac{1}{\sqrt{l}}e^{ikX_1} v_n \left( \frac{X_2}{l}
+kl \right)  \; ,
\label{eq2.3}
\end{equation} 
where $v_n (x)$ is the appropriate harmonic oscillator wave function. Here
$l=(eB)^{-1/2}$ is the magnetic length of the system and is also
the classical cyclotron radius in the lowest LL ($n=0$). 
Each level is infinitely degenerate with a degeneracy $\rho_l
=1/2\pi l^2$ per unit area.

The single particle Green function $G_0 (x,x^\prime)$ for the pure
system (disorderless) can be obtained by solving 
\begin{equation}
(\delta_\tau +H_0 -\mu)G_0 (x,x^\prime) =\delta^{(3)} (x-x^\prime)
\label{eq2.4}
\end{equation}
subject to the requirement of antiperiodicity under the translation
$\tau \rightarrow  \tau +\beta$. 
$G_0 (x,x^\prime)$
can be expanded in the normal modes of discrete frequencies
$\xi_s =(2s+1)\pi /\beta$; $s \varepsilon Z$, as
\begin{equation}
G_0 (x,x^\prime) = -\frac{1}{\beta} \sum_s e^{-i\xi_s (\tau
-\tau^\prime)} \langle X | G_0 (i\xi_s) |X^\prime \rangle  \; ,
\label{eq2.5}
\end{equation}
where the operator $G_0 (i\xi_s)$ is given by 
\begin{equation}
G_0 (i\xi_s) = \frac{1}{i\xi_s +\mu -H_0}  \;  .
\label{eq2.6}
\end{equation}
with eigen value $G_0^n (i\xi_s) =[i\xi_s +\mu -\epsilon_n
]^{-1}$. By analytical continuation $i\xi_s \rightarrow  \epsilon 
\pm i\eta $, we obtain this eigen value in the real space as
\begin{equation}
G_0^n (\epsilon ) = \frac{1}{\epsilon +\mu -\epsilon_n \pm i\eta }\;
. \label{eq2.7}
\end{equation}
Here $+(-)$ refers to retarded(advanced) Green's function.

The density of the particles is obtained in terms of the Green's
function as
\begin{equation}
\rho (X) = -G_0 (x,x^\prime)|_{X^\prime =X,\, \tau^\prime = \tau
+0} \; .
\label{eq2.8}
\end{equation}
The density is independent of the spatial coordinate, i.e., the
density is uniform, and is given by (see Ref. \cite{randj} for
details) 
\begin{equation}
p \equiv \frac{\rho}{\rho_l} = \sum_n f_n  \; ,
\label{eq2.9}
\end{equation}
where the Fermi function corresponding to $n$-th LL, $f_n =[1+ \exp
(\beta [\epsilon_n -\mu])]^{-1}$ and $p$ is the number of fully
filled LL at $T=0$.

 At low temperatures characterized by $\beta \omega_c \gg 1$, we
 shall determine the chemical potential for fixed density. 
 To this end, we write \cite{randj}
 \begin{equation}
 e^{\beta (\epsilon_n -\mu) }= w^{\mu /\omega_c -1/2 -n} \; ,
 \label{eq2.10}
 \end{equation}
 where $w = \exp [-\beta \omega_c ]$ which is the perturbative
 expansion parameter. We expand the right-hand side of
 Eq.~(\ref{eq2.9}) to obtain
 \begin{equation}
 \sum_n (1+w^{\mu /\omega_c -1/2 -n})^{-1} = \left[ 
 \frac{\mu}{\omega_c} +\frac{1}{2} \right] +
 \sum_{r \geq 1} (-1)^r \frac{w^{r\delta} -w^{r(1-\delta)} -w^{r(\mu
 / \omega_c +1/2)}}{1-w^r}  \; ,
 \label{eq2.11}
 \end{equation}
 where $[\mu /\omega_c +1/2]$ denotes the (positive) integer part:
 \begin{equation}
 \frac{\mu}{\omega_c}+\frac{1}{2} =
 \left[ \frac{\mu}{\omega_c}+\frac{1}{2} \right] 
 +\delta\; , \; 0 < \delta <1 \;
 . \label{eq2.12}
 \end{equation}
 Since $w$ is very small, Eq.~(\ref{eq2.9}) takes the form
 \begin{equation}
 p = \left[ \frac{\mu}{\omega_c} +\frac{1}{2} \right] 
 -w^\delta +w^{1-\delta}
 +w^{[\mu /\omega_c +1/2] +\delta} +\cdots  \; .
 \label{eq2.13}
 \end{equation}
 The solution of (\ref{eq2.13}) in the leading order is obtained as
 \begin{equation}
 \left[ \frac{\mu }{\omega_c}+\frac{1}{2} \right] =p 
 \; ,\; \delta = \frac{1}{2}+ \frac{w^p}{2\ln w} \; .
 \end{equation}
 Therefore, the chemical potential is given by
 \begin{equation}
 \mu = \frac{2\pi \rho}{m^\ast} -\frac{1}{2\beta}e^{-p\beta\omega_c}
 +\cdots \; .
 \label{eq2.15}
 \end{equation}
 On substitution of the value of $\mu$ on Eq.~(\ref{eq2.10}) we
 obtain
 \begin{equation}
 e^{\beta (\epsilon_n -\mu) }= \exp \left[ \beta\omega_c (n-p+1/2)
 \right] \left( 1+\frac{1}{2}e^{-\beta\omega_c (p+1/2) }+\cdots \right)
 \; .  \label{eq2.16}
 \end{equation}

\subsection{Impurity averaged Green's function}

 We assume a random distribution of white noise disorder
 potential $U(X)$ all over the space with probability
 distribution
 \begin{equation}
 P[U] = {\cal N} \exp \left[ -\frac{1}{2 \lambda^2} \int d^2X
 |U(X)|^2 \right] \; ,
 \label{eq2.17}
 \end{equation}
 where ${\cal N}$ is the renormalization constant. The average
 potential 
 \begin{equation}
 \overline{U(X)} =0  \; .
 \label{eq2.18}
 \end{equation}
 Hereafter overbar denotes the average with respect to the
 distribution $P[U]$ (Eq.~(\ref{eq2.17})) of the disorder
 potential. 
 The correlation between the potentials at two points
 is given by 
 \begin{equation}
 \overline{U(X)U(X^\prime)} = \lambda^2 \delta (X-X^\prime) \; .
 \label{eq2.19}
 \end{equation}
 The scattering potentials are short ranged.

 The single particle Green's function for the disordered system in
 terms of the free Green's function (\ref{eq2.5}) is given by
 \begin{eqnarray}
 G(x,x^\prime) &\equiv& \langle x | G |  x^\prime \rangle 
 = \langle x | G_0 +G_0 UG |x^\prime \rangle \nonumber \\
 &=& \langle x| G_0 | x^\prime \rangle + \langle x | G_0 U G_0
 |x^\prime \rangle + \langle x | G_0 UG_0UG_0 |x^\prime \rangle
 +\cdots \; .
 \label{eq2.20}
 \end{eqnarray}
 When we average over the distribution in disorder, the terms
 with odd number of $U$ operators vanish. Further, due to
 uncorrelated potentials (\ref{eq2.19}), each scattering centre
 is responsible for an even number of scatterings.

 We consider here a weak disorder, and assume that $U/\omega_c \ll
 1$ \cite{halp}. 
 Further, the parameter $\lambda \ll 1$. 
 We may thus neglect the contribution of the
 off-diagonal matrix elements 
 $\langle m \left\vert U
 \right\vert n \rangle$ and consider only the diagonal terms
 $\langle n \left\vert U \right\vert n \rangle$ in the
 determination of $G$. As a consequence, the full Green's function
 given in Eqn(3.20) will also be diagonal in the Landau level
 basis.
 If $G_0^n (\epsilon)$ be the free Green's function for $n$-th LL,
 then the corresponding impurity averaged Green's function
 is given by
 \begin{equation}
 \overline{G^n (\epsilon)} = G_0^n (\epsilon) +
 G_0^n (\epsilon)\Sigma_n (\epsilon)
 \overline{G^n (\epsilon)} \; .
 \label{eq2.21}
 \end{equation}
 In other words,
 \begin{equation}
 \overline{G^n (\epsilon)} = \frac{1}{\epsilon +\mu -\epsilon_n
 -\Sigma_n (\epsilon)} \; .
 \label{eq2.22}
 \end{equation}
 $\overline{G^n (\epsilon)}$ has the diagramatic representation
 as in Fig.~1(a). Here $\Sigma_n$ is
 the corresponding self energy. For the evaluation of $\Sigma_n
 (\epsilon)$, we make the self consistent Born approximation
 (SCBA). In SCBA, one neglects multiple scattering at any point
 and the cross scatterings which are shown in Fig.~1(b). The
 diagrams which contribute to SCBA are shown in Fig.~1(c).

 We find that $\Sigma_n$ (see Fig.~1(a)) is 
 \begin{equation}
 \Sigma_n (\epsilon) = \lambda^2 \rho_l \sum_{m \geq -n}
 \frac{1}{\epsilon +\mu -\epsilon_{n+m}-\Sigma_{n+m} (\epsilon)} \;
 . \label{eq2.23}
 \end{equation}
 The solution of Eq.~(\ref{eq2.23}) can be obtained self
 consistently \cite{pruis2}. This is given by
 \begin{equation}
 \Sigma_n (\epsilon) = -\frac{1}{2}(\epsilon +\mu -\epsilon_n)
 +\frac{1}{2}\sqrt{(\epsilon +\mu -\epsilon_n)^2 -\Gamma^2} \; ,
 \label{eq2.24}
 \end{equation}
 where $\Gamma^2 =4 \lambda^2 \rho_l$.

 We now make a few remarks \cite{referee} on SCBA and its
 limitations. In SCBA, the Landau levels are broadened but the
 broadening is very small compared to Landau level spacings,
 i.e., $\Gamma /\omega_c \ll 1$. This is, in fact, an outcome of
 the assumption that the magnetic field is very high, and that
 one has a weak disorder potential caused by a small impurity
 concentration.
 The contributions of multiple and cross scatterings to the electron
 self energy are negligible as they are of higher order in impurity
 concentration \cite{mahan}. The SCBA does not lead to any
 localized state. As a consequence, it cannot describe
 localization-delocalization transitions between two successive
 IQHE states. Considering all the scattering
 diagrams as well as the impurity induced mixing of Landau
 levels, a more  rigourous analysis would 
 presumably give rise to localized states and 
 explain the critical behaviour of
 localization-delocalization transitions. The scaling law, which
 was derived field theoretically earlier by Pruisken
 \cite{pruisken}, in critical transitions between the plateaus
 has, in fact,  been obtained by Huo, Hetzel, and Bhatt \cite{bhatt}
 in an exact
 numerical simulation. However, we remark that  SCBA is good 
 enough for obtaining
 the transport properties at integer filling factors, in
 particular.  Further, as Pruisken \cite{pruis2} has shown, 
 it can be improved
 by a renormalization group analysis to understand the formation
 of plateaus.
 This point is elaborated upon somewhat more in section IV-A.

 For a disorderless system, $\Gamma =0$ and hence the self energy
 $\Sigma_n$ also vanishes. In the range of energy $\epsilon$ for
 which $|\epsilon +\mu -\epsilon_n | >\Gamma$, $\Sigma_n$ is real;
 it merely changes the energy of the particles. On the other
 hand, for the range $|\epsilon +\mu -\epsilon_n |<\Gamma$,
 $\Sigma_n$ has both real and imaginary parts, and is given by
 \begin{equation}
 \Sigma_n^\pm (\epsilon) = -\frac{1}{2} ( \epsilon +\mu -
 \epsilon_n)\pm \frac{i}{2}\sqrt{\Gamma^2 -(\epsilon +\mu
 -\epsilon_n)^2} \; .
 \label{eq2.25}
 \end{equation} 
 Therefore in this range, the impurity averaged retarded and
 advanced Green's functions are given by
 \begin{equation}
 \overline{G^n_{r,a}(\epsilon )} = \frac{2}{\epsilon +\mu
 -\epsilon_n \pm i \sqrt{(\epsilon +\mu -\epsilon_n)^2
 -\Gamma^2}}  \; .
 \label{eq2.26}
 \end{equation}
 The DOS for the $n$-th Landau band,
 \begin{eqnarray}
 {\cal D}_n (\epsilon ) &\equiv& -\frac{\rho_l}{\pi} \, {\rm Im}\,
 \overline{G^n_r (\epsilon )}  \nonumber \\
 &=& \frac{1}{2\pi l^2}\frac{2}{\pi \Gamma}\left[ 1- \left(
 \frac{\epsilon +\mu -\epsilon_n}{\Gamma} \right)^2
 \right]^{1/2}  \; . 
 \label{eq2.27}
 \end{eqnarray}
 This is the well known SCBA DOS which is semi
 circular in nature. This will be used to determine the linear
 response of the system.

 Alternatively, one may also consider the Gaussian DOS
 which is given by
 \begin{equation}
 {\cal D}_n (\epsilon )= \frac{1}{2\pi l^2} \sqrt{\frac{2}{\pi
 \Gamma}} e^{-2(\epsilon +\mu -\epsilon_n)^2/\Gamma^2 }\;
 . \label{eq2.28}
 \end{equation}
 A brief comparison between the above two DOS will be made later.

\section{RESPONSE FUNCTION}

The partition function corresponding to the external 
electromagnetic probe $A_\mu$ is
\begin{equation}
{\cal Z}_{I} [A_\mu ] =\int [dA_\tau^{{\rm in}}] \exp [-S^E]
\; , \label{eq3.1}
\end{equation}
where $S^E$ is identified as the Euclidean one-loop effective 
action of the gauge fields. It is given by 
\begin{eqnarray}
S^E &=& \frac{1}{2} \int d^3x\int d^3x^\prime \left( A_\mu (x)
	  +A_\tau^{{\rm in}}\delta_{\mu \tau} (x) \right) 
	   \Pi_{\mu\nu}^E (x,\, x^\prime) \left(
   A_\nu (x^\prime) +A_\tau^{{\rm in}}\delta_{\nu\tau}
   (x^\prime) \right) \nonumber \\
   & & -\frac{1}{2} \int d^3x \int d^3x^\prime A_\tau^{{\rm in}}
     (x)V^{-1}(x-x^\prime) A_\tau^{{\rm in}} (x^\prime) \; .
     \label{eq3.2}
\end{eqnarray}
Here the integration over the imaginary time $\tau$ runs
from $0$ to $\beta$. The effective action is obtained by  
expanding the fermionic determinant upto quadratic terms
in powers of fields $A_\mu$ and $A_\tau^{{\rm in}}$. The 
polarization tensor $\Pi_{\mu\nu}^E (x,\, x^\prime)$ which
is evaluated at the saddle point mentioned above is 
impurity averaged. Since $\Pi_{\mu\nu}^E$ is evaluated in
the path integral formalism, $\tau $-ordering is implied
for the correlations, i.e.,
\begin{equation}
\Pi_{\mu\nu}^E (x,\, x^\prime) = - \overline{ \left \langle 
T_\tau j_\mu (x)j_\nu (x^\prime) \right \rangle } -
\overline{ \left \langle \frac{\delta j_\mu (x)}{\delta A_\nu 
(x^\prime)} \right \rangle } \; .
\label{eq3.3}
\end{equation}
The overbar implies the average over the distribution of
random disorder in Eq.~(\ref{eq2.17}). 
In terms of the thermal single particle Green's function,
the components of $\Pi_{\mu\nu}^E$ are given by
\begin{mathletters}
\begin{eqnarray}
\label{eq3.4}
\Pi^E_{\tau \tau}(x,x^\prime) &=& -e^2 \overline{G(x,x^\prime)
		 G(x^\prime ,x) }   \; ,  \\
\Pi^E_{k\tau}(x,x^\prime) &=& \frac{e^2}{2m^\ast}\left[ \overline{
     D_k G(x,x^\prime) G(x^\prime ,x) } - \overline{G(x,x^\prime)
                 D_k^\ast G(x^\prime ,x) } \right] \; ,  \\
\Pi^E_{kl}(x,x^\prime) &=& -\frac{e^2}{4m^{\ast 2}}\left[ 
   \overline{ D_k G(x,x^\prime )D_l^\prime G(x^\prime ,x)}
              -\overline{ \left( D_kD_l^{\prime\ast}G(x,x^\prime)\right) 
			G(x^\prime ,x)} \right. \nonumber \\
              & & +\overline{ D_l^{\prime\ast}G(x,x^\prime)
		  D_k^\ast G(x^\prime ,x) }
	      - \left. \overline{ G(x,x^\prime)D_k^\ast
                   D_l^\prime G(x^\prime ,x)} \right] \nonumber \\
	      & & +\frac{e^2}{m^\ast}\delta_{kl}
	      \delta (x-x^\prime) \overline{G(x,x^\prime)}
		\Big\vert_{X^\prime =X\, ,
\, \tau^\prime =\tau +0^+} \; .                                        
\end{eqnarray}
\end{mathletters}

We remark here that
once we average over the distribution of the impurity potential,
the translational invariance gets restored. Therefore, the
polarization tensor can be represented by Fourier expansion 
\begin{equation}
\Pi_{\mu\nu}^E (x, x^\prime) =\frac{1}{\beta} \sum_j \int
\frac{d^2{\bf q}}{(2\pi)^2} e^{-i{\bf q}\cdot (X-X^\prime)}
e^{-i\omega_j (\tau -\tau^\prime) }\Pi_{\mu\nu}^E (\omega_j , 
{\bf q})  \; ,
\label{eq3.5}
\end{equation}
where the Matsubara frequencies are given by
\begin{equation}
\omega_j =\frac{2\pi}{\beta}j \; ; \; j \varepsilon Z \; .
\label{eq3.6}
\end{equation}
Again, due to rotational symmetry and gauge invariance, there 
are three independent form factors. One thus obtains the
polarization tensor in the form (in momentum space)
\begin{eqnarray}
\Pi_{\mu\nu}^E (\omega_j , {\bf q}) &=& \Pi_0 (\omega_j, {\bf q})
(q^2 \delta_{\mu\nu} -q_\mu q_\nu) +(\Pi_2 -\Pi_0)({\bf q}^2
\delta_{ij}-q_i q_j) (\omega_j, {\bf q}) \nonumber \\
& & \times \delta_{\mu i}\delta_{\nu j} +\Pi_1(\omega_j, {\bf q})
\varepsilon_{\mu\nu \lambda }q^\lambda \; ,
\label{eq3.7}
\end{eqnarray}
where $q^2 =\omega_j^2 +{\bf q}^2$. Here $\Pi_0$, $\Pi_1$ and $\Pi_2$
are the form factors. Note that $\Pi_1$ is a ${\cal P}$, ${\cal T}$
violating form factor. The one loop effective action then 
acquires the form
\begin{eqnarray}
S_{{\rm effective}}^E &=& \frac{1}{2\beta} \sum_j \int \frac{d^2
{\bf q}}{(2\pi)^2} \left[ \left( A_\mu (q) +A_\tau^{{\rm in}}(q)
\delta_{\mu \tau}\right) \Pi_{\mu\nu}^E (\omega_j, {\bf q}) 
\left( A_\nu (-q) +A_\tau^{{\rm in}}(-q) \delta_{\nu \tau} \right)
\right.  \nonumber \\
& & - \left. A_\tau^{{\rm in}}(q) V^{-1} ({\bf q}) A_\tau^{{\rm in}}
\right]   \; .
\label{eq3.8}
\end{eqnarray}

\subsection{Off-diagonal Conductivity}

A straight forward linear response analysis from
Eqs.~(\ref{eq3.7}) and (\ref{eq3.8}), yields the expression
for off-diagonal conductivity to be
\begin{equation}
\sigma_{xy} = \lim_{\omega \rightarrow  0} \, {\rm Re }\,  \Pi_1
(\omega ,0) \; ,
\label{eq3.9}
\end{equation}
subject to the condition $\lim_{{\bf q} \rightarrow  0} V({\bf
q}){\bf q}^2 =0$. $\Pi_1 (\omega ,0)$ is obtained by the
analytical continuation: $i\omega_j \rightarrow  \omega +i\delta$.

The form factor $\Pi_1$ can be determined from the evaluation of the
component, say $\Pi_{1\tau}^E$ of the polarization tensor
(\ref{eq3.7}). The impurity average over two Green's functions in
Eq.~(\ref{eq3.4}) is not equal to the product of two averaged
Green's functions (Fig.~2(a)), in general. 
It has an additional contribution coming from vertex corrections
(see Fig.~2(b)). We shall determine the contributions from Figs.~2(a)
and 2(b) separately.

Considering figure 2(a), we get
\begin{equation}
\Pi_1^{(0)} (\omega_j ,0) =\frac{e^2}{2m^\ast}\frac{1}{2\pi l^2}
\sum_{n=0}^\infty \sum_{m=0}^\infty I_{nm}(\omega_j)
[(n+1)\delta_{m,n+1} +n \delta_{m,n-1} -(2n+1)\delta_{mn} ] \; ,
\label{eq3.10}
\end{equation}
where 
\begin{equation}
I_{nm}(\omega_j) = -\frac{1}{\beta} \sum_{s=-\infty}^\infty
\overline{G^n (i\xi_s)}\;\; \overline{G^m (i\xi_s -i\omega_j)}  \; .
\label{eq3.11}
\end{equation}
The evaluation of $I_{nm}$ is done in the Appendix A. 
Performing analytical
continuation: $i\omega_j \rightarrow  \omega +i\delta$, we obtain
the real part of $I_{nm}$ (see Eq.~(\ref{eqa9})),
\begin{eqnarray}
{\rm Re} \;I_{nm}(\omega) &=&
2\int_{\epsilon_n -\mu -\Gamma}^{\epsilon_n -\mu +\Gamma} 
\frac{d \epsilon }{2\pi} n_F(\epsilon ) \; {\rm Im} \; \overline{G^n_r
(\epsilon )} \;  {\rm Re} \; \overline{G^m_a(\epsilon -\omega)} 
\nonumber \\ & &
+2\int_{\epsilon_m -\mu -\Gamma}^{\epsilon_m -\mu +\Gamma} 
\frac{d \epsilon }{2\pi} n_F(\epsilon ) \; {\rm Im} \; \overline{G^m_r
(\epsilon )} \; {\rm Re }\; \overline{G^n_r(\epsilon +\omega)} \; .
\label{eq3.12}
\end{eqnarray} 
Here $n_f (\epsilon )$ represents the Fermi function,
\begin{equation}
n_F (\epsilon ) = \frac{1}{1+e^{\beta \epsilon }}  \; .
\label{eq3.13}
\end{equation}

By Taylor expansion of Eq.~(\ref{eq3.12}) in powers of $\Gamma
/\omega_c$, we obtain, with the use of Eqs.~(\ref{eq3.9}) --
(\ref{eq3.13}), the off-diagonal conductivity (without vertex
correction)
\begin{equation}
\sigma_{xy}^{(0)} = -\frac{e^2}{2\pi} \left[ \sum_{n=0}^\infty f_n
-\frac{1}{2} \beta\omega_c \sum_{n=0}^\infty (2n+1)f_n (1-f_n)
\right] 
+{\cal O} \left( \frac{\Gamma}{\omega_c} \right)^2 \; .
\label{eq3.14}
\end{equation}

{\it Vertex correction:} We next consider the
contribution to $\sigma_{xy}$ due to vertex
corrections shown in Fig.~2(b). For the lowest order vertex
correction, the contribution $\sigma_{xy}^{(1)}$ is suppressed by a
factor $(\Gamma /\omega_c)^2$ in the limit $\Gamma /\omega_c \ll
1$, since two more scatterings take place in the virtual process.
The vertex corrections due to more and more number of scatterings
lead to the contributions which are further suppressed by a
factor $\Gamma /\omega_c $ for each scattering. Therefore,
in the limit $\Gamma /\omega_c \ll 1$, the total off-diagonal
conductivity is essentially given by
\begin{eqnarray}
\sigma_{xy} &\simeq & \sigma_{xy}^{(0)}  \nonumber \\
 &=& -\frac{e^2}{2\pi} \left[ \sum_{n=0}^\infty f_n
-\frac{1}{2} \beta\omega_c \sum_{n=0}^\infty (2n+1)f_n (1-f_n)
\right] 
+{\cal O} \left( \frac{\Gamma}{\omega_c} \right)^2 \; .
\label{eq3.15}
\end{eqnarray}
In agreement with Ando et al \cite{ando}, the impurity contribution
to $\sigma_{xy}$ is of the order of $(\Gamma /\omega_c)^2$ which is
small in the strong field and small broadening approximation.

Starting with the result of Ando et al \cite{ando} which was
obtained at $T=0$ using a simple Lorentzian DOS, Pruisken
\cite{pruis2} has shown that after successive renormalizations,
there is a renormalization group flow towards $(\sigma_{xx} ,
\sigma_{xy}) = (0, {\rm integer})\frac{e^2}{2\pi}$, corresponding
to the widening of the plateaus. He finds another flow towards an
unstable fixed point which corresponds to $\sigma_{xy} =(p+1/2)
\frac{e^2}{2\pi}$ where the transition between the plateaus occur.
The group flows have been observed experimentally \cite{pruis3}.
Thus at $T=0$, the stable group flow is towards the point for which
filling factor $\nu$ is an integer and $(\sigma_{xx} ,
\sigma_{xy}) = (0,\nu)\frac{e^2}{2\pi}$ which is the value of
$(\sigma_{xx} , \sigma_{xy})$ at $\Gamma =0$. Recall that we are
considering those specific values of $B$ (or $\rho$) for which $\nu
=p$ is an integer. We have computed here $\sigma_{xy}$ more
rigorously within SCBA and also at $ T \neq 0$. It is therefore
reasonable to expect the present improved treatment to be capable of
explaining the behaviour of plateaus, {\em a la} the Pruisken
analysis. It would indeed be interesting to examine whether the
stable renormalization group flow at finite $T$ occurs towards the
point $(\sigma_{xx} ,\sigma_{xy})$, which we evaluate here, for
which $\nu$ is an integer and $\Gamma =0$. This analysis would
explain the behaviour of plateau at finite temperatures in
$\sigma_{xy}$ given by Eq.~(\ref{eq3.15}). We hope to take this up
in future.

Interestingly the contribution to $\sigma_{xy}$
in Eq.~(\ref{eq3.15}) is {\it entirely} due to the temperature
in the leading order,
surviving even in the absence of impurities. At this juncture, we
recall the lore \cite{prange2} that a pure system may not be
expected to show such a behaviour, in consequence of translation
invariance. This classical argument clearly does not hold here; and
the violation may be attributed to the fact that in the presence of an
external magnetic field, the generators of translation group are
anomalous. The co-ordinates of the centre of the classical orbit do
not commute --- a feature which was labeled by Chen, Wilczek,
Witten, and Halperin \cite{chen} as {\it violation of fact}. We
note here that the formalism does {\it not} violate translation
invariance. Rather, it is analogous to the well-known field
theoretic anomalies. 
 Note that the expression for $\sigma_{xy}$ (\ref{eq3.15})
is manifestly consistent with translation
invariance. We may point out that such a temperature evolution has
been noticed earlier in the finite temperature studies 
\cite{randj,ftemp} of
Chern-Simons superconductivity (CSS). 
In fact, Fradkin \cite{frad} has in his study of CSS drawn explicit
analogy with the anomalous nature of the generators of the
translation group with the anomaly arising in the
Schwinger-Anderson model \cite{sch,ander}. Should this analogy indeed hold
perfectly, we may then look upon quantum Hall systems to provide a
condensed matter laboratory to probe the new anomaly as much as the
process $\Pi^0 \rightarrow  2\gamma $ provides, in order to probe
the chiral anomaly. We also mention that Bellissard et al
\cite{bellis} also obtain a temperature dependence for a pure
system, but they miss the crucial prefactor which is derived here.
Finally, it may be noted that this `anomalous' contribution owes
its origin to intra LL transitions, which would contribute in the
thermodynamic limit only if degeneracy grows with area.

It is interesting that the above mentioned intra Landau
level transition
which is unique to this case also contributes to the specific heat,
as derived by Zawadzki and Lassnig \cite{zawa}. This has
been experimentally verified later by Gornik et al \cite{gornik}.

At low temperatures $(\beta \omega_c \gg 1)$, we perturbatively
evaluate $\sigma_{xy}$ with the expansion parameter
$e^{-\beta\omega_c}$, as we have discussed in the previous section.
We thus find
\begin{equation}
\sigma_{xy} (T) = -\frac{e^2p}{2\pi} [1-4y] \; ,
\label{eq3.16}
\end{equation} with 
\begin{equation} y = \frac{T_0}{T}\;\exp\;\left[ -\frac{T_0 }{T} \right] 
\; , \label{eq3.17} 
\end{equation} 
where $T_0 =\omega_c /2$. 
The temperature correction to the $\sigma_{xy}$ is exponentially 
weak. This leads to some interesting 
consequences as we shall show in the next section. 

\subsection{Diagonal Conductivity}

Again, a linear response analysis of Eq.~(\ref{eq3.8}) provides the
diagonal conductivity to be 
\begin{equation}
\sigma_{xx} = - \lim_{\omega \rightarrow  0} \, {\rm Im } \,
\Pi_{11}^r (\omega ,0)  \; .
\label{eq3.18}
\end{equation}
Here $r$ represents
the retarded part of the correlation function which is obtained from
Eq.~(\ref{eq3.7}) by analytical continuation.
Equation (\ref{eq3.18}) is derived in the limit 
$\lim_{{\bf q} \rightarrow  0} V({\bf q}){\bf q}^2 =0$. Therefore
Eq.~(\ref{eq3.18}) is true for all those potentials which are short
ranged compared to $\ln r$.

We evaluate $\Pi_{11}^E (\omega_j ,0)$ using the same procedure
employed in determining $\Pi_{1\tau}^E$ discussed above.
It may again be shown that vertex corrections are suppressed by
an extra factor of $\Gamma /\omega_c$. We therefore
only need to evaluate the
contribution arising from Fig.~2(a). We thus obtain
\begin{eqnarray} 
\Pi_{11}^E (\omega_j ,0) &=& 
\frac{e^2}{4m^{\ast 2}}\frac{1}{2\pi l^4}
\sum_{n=0}^\infty \sum_{m=0}^\infty I_{nm}(\omega_j)
[(n+1)\delta_{m,n+1} +n \delta_{m,n-1} -(2n+1)\delta_{mn} ]
\nonumber \\
& & -\frac{e^2}{m^\ast} \frac{1}{2\pi l^2} \sum_{n=0}^\infty 
I_n \; ,
\label{eq3.19}
\end{eqnarray} where \begin{equation}
I_n = -\frac{1}{\beta} \sum_{s=-\infty}^\infty \overline{G^n
(i\xi_s)}  \; .
\label{eq3.20}
\end{equation}
The term $I_n$ is independent of $\omega_j$; it does not have any
imaginary part. Using Eq.(\ref{eqa9}), we write
\begin{eqnarray}
{\rm Im}\;I_{nm}(\omega) &=&
-2\int_{\epsilon_n -\mu -\Gamma}^{\epsilon_n -\mu +\Gamma} 
\frac{d \epsilon }{2\pi} n_F(\epsilon ) \;{\rm Im}\; \overline{G^n_r
(\epsilon )} \; {\rm Im}\; \overline{G^m_a(\epsilon -\omega)} 
\nonumber \\ & &
+2\int_{\epsilon_m -\mu -\Gamma}^{\epsilon_m -\mu +\Gamma} 
\frac{d \epsilon }{2\pi} n_F(\epsilon ) \;{\rm Im} \; \overline{G^m_r
(\epsilon )}\; {\rm Im}\;  \overline{G^n_r(\epsilon +\omega)} \; .
\label{eq3.21}
\end{eqnarray}

The diagonal conductivity is thus given
by 
\begin{equation}
\sigma_{xx} = -\lim_{\omega \rightarrow  0}
\frac{e^2}{4m^{\ast 2}} \frac{1}{2\pi l^4} \sum_{n=0}^\infty
\sum_{m=0}^\infty {\rm Im} \, I_{nm} (\omega)\, \left[
(n+1)\delta_{m,n+1} +n \delta_{m,n-1} -(2n+1) \delta_{mn} \right]
\; . \label{eq3.22}
\end{equation}
In the limit $\Gamma \rightarrow  0$
(disorder free), $\sigma_{xx}$ vanishes, as may easily be
checked. For $\Gamma$ finite, there are
some interesting consequences. Recall that according to SCBA,
the imaginary part of $\overline{G^n_r (\epsilon )}$ exists in the
region of $\epsilon $ for which $\Gamma > |\epsilon_n +\epsilon
-\mu |$. Therefore the integrals in Eq.~(3.21) are non-zero only
when $m=n$, i.e., the contribution is entirely due to transitions
within the band of an LL. We obtain
\begin{equation}
\sigma_{xx} = \frac{2e^2}{3\pi^2} \left(
\frac{\omega_c}{\Gamma} \right) \beta\omega_c \sum_{n=0}^\infty
(2n+1)f_n (1-f_n) +{\cal O} \left( \frac{\Gamma}{\omega_c} \right)
\; .  \label{eq3.23}
\end{equation}
Notice that at finite temperatures, there is a singular
contribution in $\sigma_{xx}$ as $\Gamma \rightarrow  0$.
However, as we have seen above, $\sigma_{xx}$ vanishes
for $\Gamma =0$ at all temperatures. Further at $T=0$, there is no
possibility of intra band transition within an LL and so
$\sigma_{xx}$ vanishes again. Therefore the expression
(\ref{eq3.23}) has a pertinent role only at finite temperatures, and
in the presence of disorder.
In short, $\sigma_{xx}(\Gamma ,T=0)=0$; $\sigma_{xx}(\Gamma
=0,T)=0$.
In the high field and low broadening
limit, $\Gamma /\omega_c \ll 1$, and so we neglect the higher order terms
in $\Gamma /\omega_c$.
Note that $\sigma_{xx} (T)$ is evaluated for those specific value
of $B$ (or $\rho$) for which $\nu =p$ is an integer.

A low temperature expansion of Eq.~(\ref{eq3.23}) yields
\begin{equation}
\sigma_{xx} (T) = \frac{16e^2p}{3\pi^2} \left(
\frac{\omega_c}{\Gamma} \right) y \equiv
\frac{16e^2p}{3\pi^2} \left( \frac{\omega_c}{\Gamma} \right) \frac{T_0}{T}
\, \exp \, \left[ -\frac{T_0}{T} \right] \; .
\label{eq3.25}
\end{equation}
It may be noted that
there is now a competition between two energy scales --- $T$ and
$\Gamma$ in the theory. For very low temperatures, $y$ is
exponentially small (see Eq.~(\ref{eq3.17})) and the value of
$\sigma_{xx}$ need not be large although $\omega_c /\Gamma \gg 1$.
In fact, at $T=0$, $y=0$ leading to a vanishing of $\sigma_{xx}$.
Further, we see that
$\sigma_{xx} (T)$ is thermally activated with a temperature
dependent prefactor, in agreement with the experiment of Katayama et
al \cite{kata}.
This  prefactor
is clearly not universal any
more, 
since it depends on the integer filling $p$ and
the broadening of an LL, as given by $\Gamma$.
The mismatch between the results of Ref.~\cite{poly}
which predicts universality and that of ours may be due to 
consideration of different
types  of the disorder potential; whereas they have taken a long
range scatterer, we have considered $\delta$-correlated short range
disorder potential. Interestingly, Fogler et al \cite{fogl} report
that for long range scatterers, the prefactor is temperature
dependent below a characteristic temperature.
In concluding this section, we note that our results
obey the phenomenological relation 
\begin{equation}
\Delta \sigma_{xy} \equiv \sigma_{xy} (T) -\sigma_{xy} (0) \sim
-\left( \frac{\Gamma}{\omega_c} \right) \sigma_{xx} \; ,
\label{eq3.26}
\end{equation}
which was proposed by Ando et al \cite{ando} based on general
arguments.

\subsubsection{$\sigma_{xx}$ for Gaussian density of states}

For the Gaussian density of states (\ref{eq2.28}), the limits of
integration in Eq.~(\ref{eq3.21}) runs from $-\infty $ to $\infty$.
There is a possibility of inter Landau band transition at finite
temperatures apart from the intra Landau band transition, since
the Gaussian density of states has a long tail, unlike the SCBA
density of states where there is a sharp cut-off. However, the
contribution to $\sigma_{xx}$ due to inter Landau band transition
is exponentially weak, and behaves like 
$\exp [-\omega_c^2 /2\Gamma^2 ]$. Note that
$\Gamma /\omega_c \ll 1$. The major contribution therefore
again comes from the
intra Landau band transition. The vertex corrections are suppressed by
the higher orders of $\Gamma /\omega_c$. We thus obtain the
diagonal conductivity for the Gaussian density of states
\begin{equation}
\sigma_{xx} \simeq \frac{e^2}{8\sqrt{2\pi}} \left(
\frac{\omega_c}{\Gamma} \right) \beta\omega_c \sum_{n=0}^\infty
(2n+1)f_n (1-f_n)  \; ,
\label{eq3.27}
\end{equation} 
which has the low temperature form
\begin{equation}
\sigma_{xx} (T) =\frac{e^2p}{\sqrt{2\pi}} \left( 
\frac{\omega_c}{\Gamma} \right) y  \; ,
\label{eq3.28}
\end{equation}
which qualitatively agrees with $\sigma_{xx} (T)$ obtained from the
SCBA density of states as in Eq.~(\ref{eq3.25}).
This exercise serves to demonstrate that the crucial features of
$\sigma_{xx}$ are of not artifacts of the SCBA.

\section{Resistivities and Comparison with Experiment}

We now take up the discussion of resistivities of quantum Hall
states which is of primary interest to us. Cage et al \cite{cage1}
and Yoshihiro et al \cite{yoshi} have found that $\rho_{xy} (T)$
decreases with increase in temperature and the deviation from its
zero temperature value, $\Delta \rho_{xy} (T) \equiv
\rho_{xy}(T)-\rho_{xy}(0)$ obeys a linear relationship with
$\rho_{xx}^{{\rm min}}$ as
\begin{equation}
\Delta \rho_{xy} =-S \rho_{xx}^{{\rm min}} \; .
\label{eq4.1}
\end{equation}
Cage et al \cite{cage1} have further reported that the linearity
remains for the temperature range $1.2$--4.2K for which the value
of $\rho_{xx}^{{\rm min}}$ changes upto four decades. The value of
$S$ depends on the device and the cooling process and hence is not
reproducible. However, the linear relationship (\ref{eq4.1}) is
universal. The sign of $S$ is {\it not} always positive as Wel et
al \cite{wel} have observed that $\rho_{xy} (T)$ increases with
temperature. We shall see that both the signs of $S$ along with the
linearity (\ref{eq4.1}) can be obtained in different physical
regimes of temperature and broadening, (although in a narrow range
of temperatures for positive $S$). Presumably, these physical
regimes correspond to the above mentioned experiments. Both the
regimes belong to low temperatures and low broadening, as we show
below.

In the previous section, we have determined $\sigma_{xy} (T)$ and
$\sigma_{xx}(T)$. Now by symmetry, $\sigma_{yx}(T) =-\sigma_{xy}
(T)$ and $\sigma_{yy}(T) =\sigma_{xx}(T)$. We then invert the
conductivity matrix to obtain the resistivity matrix with the
components
\begin{equation}
\rho_{xy}(T)
=-\frac{\sigma_{xy}(T)}{\sigma_{xx}^2(T)+\sigma_{xy}^2(T)} 
=-\rho_{yx} (T) \; ;\;
\rho_{xx}(T) 
=\frac{\sigma_{xx}(T)}{\sigma_{xx}^2(T)+\sigma_{xy}^2(T)} 
=\rho_{yy} (T) \; .
\label{eq4.2}
\end{equation}
Using the conductivities obtained from SCBA DOS
(Eqs.~\ref{eq3.16} and \ref{eq3.25}), we get
\begin{mathletters}
\label{eq4.3}
\begin{eqnarray}
\rho_{xy}(T) &=& \left( \frac{2\pi}{e^2p} \right)
\frac{1-4y}{(32/3\pi)^2(\omega_c /\Gamma)^2 y^2 +(1-4y)^2} \\
\rho_{xx}(T) &=& \left( \frac{2\pi}{e^2p} \right)
\frac{(32/3\pi)(\omega_c /\Gamma) y}
{(32/3\pi)^2(\omega_c /\Gamma)^2 y^2 +(1-4y)^2}  \; .
\label{eq4.3b}
\end{eqnarray}
\end{mathletters}
We shall study these resistivities for two different 
situations below.

(i) If  $\Gamma /\omega_c$ is so small that $(\Gamma /\omega_c)^2 =
\alpha y$ with $\alpha \sim 1$, then $(\omega_c /\Gamma)^2 y^2 \sim
y$. We then obtain
\begin{mathletters}
\label{eq4.4}
\begin{eqnarray}
\rho_{xy} (T) &=& \rho_{xy} (0) \left[ 1- \left\{ \alpha \left(
\frac{32}{3\pi}\right)^2 -4\right\}y \right] \\
\rho_{xx}(T) &=& \frac{2\pi}{e^2p} \left( \frac{32}{3\pi}\right)
\left[ \frac{\omega_c}{\Gamma} \right] y  \; ,
\label{eq4.4b}
\end{eqnarray}
\end{mathletters}
where $\rho_{xy}(0) = 2\pi /e^2p$ is the Hall resistivity at $T=0$.

First of all, $\rho_{xx}(T)$ which is thermally activated
with the activation gap
$T_0 =\omega_c /2$ with a prefactor $\sim 1/T$, is in agreement with the
best fit data of Weiss et al \cite{weiss}, who find that in the
lowest order of $y$, $\rho_{xx}^{{\rm min}} \sim (1/T) \exp
[-\omega_c /2T]$. 
Note that since we are at the centre of the plateau ($\nu$ is an
integer), Eqs.~(\ref{eq4.3b}) and (\ref{eq4.4b}) do indeed
correspond to the $\rho_{xx}^{{\rm min}}$ which is measured. 
Secondly, the Hall resistivity {\it decreases} with increase in
temperature, with $\Delta \rho_{xy} (T)$ also being thermally
activated, again in agreement with experiments of Cage et al
\cite{cage1} and Yoshihiro et al \cite{yoshi}. Finally, the linear
relation (\ref{eq4.1}) for $S>0$ holds in the rather narrow range
of temperatures over
which $(\Gamma /\omega_c)^2 \sim y$. In such a case $S$ (which is
effectively temperature independent) is given by
\begin{equation}
S = \frac{3\pi}{32}\left[ \alpha \left( \frac{32}{3\pi}\right)^2 -4
\right] \left( \frac{\Gamma}{\omega_c} \right) \; .
\label{eq4.5}
\end{equation}
The constancy of $S$ remains over a decade in $\rho_{xx}$, which is
smaller than that of cage et al \cite{cage1} who find $S$ to be
constant over four decades in $\rho_{xx}$.

(ii) Consider the case where $\Gamma /\omega_c$ is small 
$(\Gamma /\omega_c \ll 1)$ but $(\Gamma /\omega_c)^2 \gg
y$. Then we obtain from Eq.~(\ref{eq4.3}), 
\begin{mathletters}
\label{eq4.6}
\begin{eqnarray}
\rho_{xy}(T) &=& \rho_{xy} (0) \left[ 1+4y \right] \\
\rho_{xx} (T) &=& \left( \frac{2\pi}{e^2p} \right) \frac{32}{3\pi}
\left( \frac{\omega_c}{\Gamma}\right) y  \; .
\end{eqnarray}
\end{mathletters}
Here $\rho_{xy} (T)$ {\it increases} with temperature
and $\rho_{xx}(T)$ behaves the same way as in case(i).
Interestingly, in this case the linear relation (\ref{eq4.1})
between $\rho_{xx}$ and $\Delta \rho_{xy}$ 
is more rigorous, with a positive slope given by
\begin{equation}
S =-\frac{3\pi}{8}\left( \frac{\Gamma}{\omega_c} \right) \; .
\label{eq4.7}
\end{equation}
Such a behaviour is seen by Wel et al \cite{wel}.

We shall now proceed to estimate the effective masses $m^\ast$
from the reported measurements \cite{yoshi,cage1} below. 
It is clear from the discussion above that we can hope only to get
overall order of magnitude agreement.
Given the accuracy of quantization
\begin{equation}
{\cal R} \equiv \left\vert \frac{\Delta \rho_{xy}}{\rho_{xy}(0)}
\right\vert 
\label{eq4.8}
\end{equation}
at temperature $T$, the carrier density $\rho$ and the filling
factor $p$, one can estimate $m^\ast$ from Eq.~(\ref{eq4.4}). We
find, for the choice $\alpha =1$ in Eq.~(\ref{eq4.4}), 
with the reported accuracy of 4.2
ppm at $T=3$K for $p=4$ IQHE states in a GaAs sample having $\rho
=5.61\times 10^{11}$ cm.$^{-2}$, $m^\ast =0.075$ m$_e$. Their
further measurement at $T=1.2$K with accuracy 0.017 ppm gives the
value of $m^\ast =0.14$ m$_e$. In the measurement of Yoshihiro et
al \cite{yoshi} in Si MOSFET with $\rho = 1.0\times 10^{12}$
cm.$^{-2}$, they find ${\cal R}= 0.2$ ppm at $T=0.5$K for $p=4$
IQHE state. We estimate $m^\ast =0.67$ m$_e$ from this measurement
for Si MOSFET sample. 
We know that the values of $m^\ast$ in GaAs and Si MOSFET samples
are 0.07m$_e$ and 0.2m$_e$ respectively.
Note that it is possible to improve the agreement by further fine
tuning the paremaeter $\alpha$. In short,
the estimated effective masses
agree with the known ones for the respective samples in the order
of magnitude.

\section{CONCLUSION}

We have considered a quantum Hall system in the presence of weak
disorder and determined the diagonal conductivity $\sigma_{xx}$ and
off-diagonal conductivity $\sigma_{xy}$ at finite temperatures,
within the SCBA. We have considered the only integer value of filling
factor $\nu$ which is in fact the central point of a given plateau
in $\rho_{xy}$ and the point of minimum in $\rho_{xx}$. We have shown
that $\sigma_{xy}(T)$ acquires a novel temperature dependence
which, in fact, sustains in the limit of vanishing impurity. This
`anomalous' dependence, which is contrary to the classical
expectation, is due to the anomalous nature of the generators
of the translation group. $\sigma_{xx}$ vanishes in either of the
zero temperature and zero impurity limits. i.e., $\sigma_{xx}$ is
non-zero only when both temperature and impurity are non-zero. At
low temperatures, $\sigma_{xx}(T)$ does activate with a temperature
dependent prefactor. Inverting the conductivity matrix, we obtain
$\rho_{xx}(T)$ and $\rho_{xy}(T)$. We have found that depending on the
physical regimes of $T$ and $\Gamma$, $\rho_{xy}(T)$ either
decreases or increases with increase in temperature. $\rho_{xx}(T)$
is thermally activated as $\rho_{xx} \sim (1/T)\, \exp \,
[-\omega_c /2T]$. Further $\Delta \rho_{xy}$ is shown to obey a
linear relationship with $\rho_{xx}$ in agreement with the
experiments. An estimation of effective mass from the measured
value of $\rho_{xy}(T)$ turns out to be reasonable, and finally,
our determination of compressibility (appendix B)
shows that there is no phase
transition involved, again in agreement with experiments.

Finally, a remark on Fractional Quantum Hall states.
For obtaining the temperature dependence of Hall and diagonal
conductivities for these states, one may
perform a similar analysis within the composite fermion model
\cite{jain}, since composite fermions fill an integer number of
Landau levels.
There is, however,  an additional complication involving the 
integration
over the Chern-Simons gauge field (which attaches an even number
of flux tubes to each particle) in the determination of the
response functions. This is planned to be taken up
in future.

\section*{ACKNOWLEDGMENTS}

We thank Avinash Singh for helpful discussions and suggestions.
We thank the referees for their useful suggestions and queries.

\appendix

\section*{A}

\subsection*{The evaluation of $I_{nm}$}

In this Appendix, we shall evaluate
\begin{equation}
I_{nm}(\omega_j) = -\frac{1}{\beta} \sum_{s=-\infty}^\infty
\overline{G^n (i\xi_s)}\; \overline{G^m (i\xi_s -i\omega_j)}  \; ,
\label{eqa1}
\end{equation}
where 
\begin{equation}
\overline{G^n (i\xi_s)} = \frac{1}{i\xi_s +\mu - \epsilon_n
-\Sigma_n (i\xi_s)}
\label{eqa2}
\end{equation}
with
\begin{equation}
\Sigma_n (i\xi_s) = \frac{1}{2}(i\xi_s +\mu -\epsilon_n) \pm
\frac{i}{2} \sqrt{\Gamma^2 -(i\xi_s +\mu -\epsilon_n)^2}  \; .
\label{eqa3}
\end{equation}
We evaluate $I_{nm}$ by the contour integration method
\cite{mahan}.

Let us consider the integral
\begin{equation}
S_{nm} = \int_C \frac{dZ}{2\pi i} n_F (Z) \overline{G^n (Z)} \;
\overline{G^m (Z-i\omega_j)}  \; ,
\label{eqa4}
\end{equation} 
where 
\begin{equation}
n_F(Z) = \frac{1}{1+e^{\beta Z}}  
\label{eqa5}
\end{equation}
is the Fermi function. The contour is shown in Fig.~3. There are
poles at $Z= i\frac{\pi}{\beta}(2s+1)$. The integrand has also
branch cuts on the two lines $Z=\epsilon$ and $Z=\epsilon
+i\omega_j$, where $\epsilon$ is real. To illustrate, for
$\overline{G^n (Z)}$, the range of $\epsilon$ is $\epsilon_n -\mu
-\Gamma$ to $\epsilon_n -\mu +\Gamma$. The evaluation of residues
at the poles of $S_{nm}$ reproduces $I_{nm}$. In other words,
$I_{nm}$ is equivalent to $S_{nm}$ when $S_{nm}$ is evaluated along
the branch cuts, where the contributions above and below the cuts
are subtracted:
\begin{eqnarray}
I_{nm} (\omega_j) 
&=& \int_{\epsilon_n -\mu -\Gamma}^{\epsilon_n
-\mu +\Gamma} \frac{d\epsilon }{2\pi i} n_F(\epsilon )
\overline{G^m (\epsilon -i\omega_j)} \; \left[
\overline{G^n(\epsilon +i\delta)}\; -\; \overline{G^n(\epsilon -i
\delta )} \right]  \nonumber \\
& & +\int_{\epsilon_m -\mu -\Gamma}^{\epsilon_m
-\mu +\Gamma} \frac{d\epsilon }{2\pi i} n_F(\epsilon )
\overline{G^n (\epsilon +i\omega_j)} \; \left[
\overline{G^m(\epsilon +i\delta)}\; -\; \overline{G^m(\epsilon -i
\delta )} \right]  \; .
\label{eqa6}
\end{eqnarray}
Now the factor
\begin{eqnarray} 
\overline{G^n(\epsilon +i\delta)}\; -\; \overline{G^n(\epsilon -i
\delta)} &=& \overline{G^n_r (\epsilon )} \; -\;
\overline{G^n_a(\epsilon)}  \nonumber \\
& =& 2i\, {\rm Im} \, \overline{G^n_r(\epsilon )} \; .
\label{eqa7}
\end{eqnarray}
Therefore
\begin{eqnarray}
I_{nm}(\omega_j) 
&=& 2\int_{\epsilon_n -\mu -\Gamma}^{\epsilon_n
-\mu +\Gamma} \frac{d\epsilon }{2\pi } n_F(\epsilon )
\overline{G^m (\epsilon -i\omega_j)} \; {\rm Im }\, \overline{G^n_r
(\epsilon )}  \nonumber \\
& & +2\int_{\epsilon_m -\mu -\Gamma}^{\epsilon_m
-\mu +\Gamma} \frac{d\epsilon }{2\pi } n_F(\epsilon )
\overline{G^n (\epsilon +i\omega_j)} \; {\rm Im} \, \overline{G^m_r
(\epsilon )}   \; .
\label{eqa8}
\end{eqnarray}

We now analytically continue: $\omega_j \rightarrow  \omega
+i\delta$ to obtain
\begin{eqnarray}
I_{nm}(\omega +i\delta) 
&=& 2\int_{\epsilon_n -\mu -\Gamma}^{\epsilon_n
-\mu +\Gamma} \frac{d\epsilon }{2\pi } n_F(\epsilon )
\; {\rm Im} \,\left[ \overline{G^n_r (\epsilon )} \right] \,  
\overline{G^m_a (\epsilon -\omega)} \nonumber \\
& & +2\int_{\epsilon_m -\mu -\Gamma}^{\epsilon_m
-\mu +\Gamma} \frac{d\epsilon }{2\pi } n_F(\epsilon )
\; {\rm Im }\, \left[ \overline{G^m_r (\epsilon )} \right] \, 
\overline{G^n_r (\epsilon +\omega)}  \; ,
\label{eqa9}
\end{eqnarray} 
which is the required result.

\section*{B}

\subsection*{Compressibility}

The compressibility of the system can be obtained from the
density-density correlation $\Pi_{00}$ as
\begin{equation}
\kappa = \left( \frac{1}{e^2\rho^2} \right) \lim_{{\bf q}
\rightarrow  0} Re \, \Pi_{00} (0, {\bf q}) \; .
\label{eq3.29}
\end{equation} 
We now evaluate $\Pi_{\tau \tau}^E (\omega_j \, ,{\bf q})$ which
reduces to $\Pi_{00}(\omega , {\bf q})$ with the analytical
continuation.

Without the vertex corrections, we find
\begin{eqnarray}
\Pi_{\tau\tau}^{(0)E} &=& -\frac{e^2}{2\pi l^2} \sum_{n=0}^\infty
\sum_{m=0}^\infty I_{nm}(\omega_j) \left[ \delta_{mn} \right.
\nonumber \\
& & +\left. \frac{{\bf q}^2l^2}{2} \left\{ (n+1)\delta_{m,n+1} +n
\delta_{m,n-1} -(2n+1)\delta_{mn} \right\} \right] +{\cal O} (({\bf
q}^2)^2)  \; .
\label{eq3.30}
\end{eqnarray}
Using the Eq.~(\ref{eqa9}), we evaluate $\kappa $ by the Taylor
expansion in powers of $\Gamma /\omega_c$ as before. The vertex
contributions will not be considered as those are in higher order
of $\Gamma /\omega_c$. We, therefore, find
\begin{equation}
\kappa = \left( \frac{1}{e^2\rho^2} \right) {\cal K} \; ,
\label{eq3.31}
\end{equation} 
where
\begin{equation}
{\cal K} = \frac{e^2 m^\ast}{2\pi} \beta\omega_c \sum_{n=0}^\infty f_n
(1-f_n) + {\cal O} \left( \frac{\Gamma}{\omega_c} \right)^2 \; .
\label{eq3.32}
\end{equation}
Again, the thermal contribution to $\kappa $ is entirely due to
intra Landau band transition.

A low temperature expansion of Eq.~(\ref{eq3.32}) yields
\begin{equation}
\kappa = \frac{2m^\ast}{\pi\rho^2} \frac{T_0}{T} \, \exp \, \left[
-\frac{T_0}{T} \right] \; .
\label{eq3.33}
\end{equation}
At T=0, $\kappa =0$, i.e., the Hall fluid is incompressible, as we
know. At low temperatures, $\kappa $ is non-zero but exponentially
small and hence the fluid is effectively incompressible. Figure 4
shows how the value of $\kappa $ increases with increase in
temperature for a given values of $\rho /m^\ast$ and $p$. The
smooth behaviour of $\kappa $ with temperature upto a value 5K
clearly shows that there is no phase transition involving the
fluids. The same conclusion has been arrived at by Chang et
al \cite{chang} by their measurement of $\rho_{xx} (T)$ for a wide
range of temperatures.

\newpage

\begin{center}

{\bf FIGURE CAPTIONS}
\end{center}

\bigskip

FIG.1 : The thick (thin) lines with arrows represent disordered
(disorderless) single particle Green's functions. The symbol X
denotes the scattering centre and the dashed lines represent the
scatterings. (a) The propagator (impurity averaged) for $n$-th LL
$\overline{G^n}$ is shown. Here $\Sigma_n$ is the corresponding
self energy. (b) The diagrams which are not taken into account in
the SCBA --- the diagrams corresponding to multiple scatterings at
a given scattering centre and the cross diagrams. (c) The diagrams
which have been considered for the evaluation of self energy.\\

FIG.2 : The thick lines with the arrows represent the single
particle Green's function. The wiggly lines represent the
external electromagnetic probe. (a) The response function
$\Pi_{\mu\nu}$ within the SCBA is shown. (b) The diagrams
corresponding to the vertex corrections for the response
function.\\

FIG.3 : The contour for $S_{nm}$ in Eq.~(\ref{eqa4}) is shown.
The wiggly lines represent the branch
cuts along the lines $Z=\epsilon$ and $Z=\epsilon +i\omega_j$. The
cross points are the poles located at
$Z=i\frac{\pi}{\beta}(2s+1)$.\\

FIG.4 : The compressibility as a function of temperature
for the integer quantum Hall state $\nu =1$. We have chosen $m^\ast
=0.07$ m$_e$ and $\rho =2\times 10^{11}$ cm.$^{-2}$.

\begin{references}
\bibitem{fnote1} Alternatively, one may also use Hamiltonian
			formalism in the frame work of Random Phase
			approximation (RPA). The Gaussian fluctuation
			in the Lagrangian formalism is equivalent to 
			RPA in Hamiltonian formalism.
\bibitem{referee} We owe these remarks to the suggestions of the
		  referees.
\end{references}
\end{document}